\documentclass[twocolumn,aps]{revtex4-1}

\usepackage{graphicx}
\usepackage{amssymb}
\usepackage{amsfonts}
\usepackage{amsmath}
\usepackage{amsbsy}
\usepackage{natbib}
\usepackage{amsmath}
\usepackage{color}
\usepackage{float}

\renewcommand{\vec}[1]{\mbox{\boldmath$\mathrm{#1}$}}
\newcommand{\be}{\begin{equation}}
\newcommand{\ee}{\end{equation}}
\newcommand{\ben}{\begin{eqnarray}}
\newcommand{\een}{\end{eqnarray}}

\def  \bdelta    {\mbox{\boldmath$\delta$}}

\begin{document}

\title{Rectification of the spin Seebeck current in noncollinear antiferromagnets}

\author{ L. Chotorlishvili$^{1,2}$, Xi-guang Wang$^3$, A. Dyrda\l$^4$, Guang-hua Guo$^3$, V.K. Dugaev$^{2}$, J. Barna\'{s}$^{4,5}$, J. Berakdar$^2$}
\address{$^1$ Department of Physics and Medical Engineering, Rzeszow University of Technology, 35-959 Rzeszow, Poland\\
$^2$ Institut f\"ur Physik, Martin-Luther Universit\"at Halle-Wittenberg, D-06120 Halle/Saale, Germany \\
$^3$ School of Physics and Electronics, Central South University, Changsha 410083, China\\
$^{4}$ Faculty of Physics, Adam Mickiewicz University, ul. Umultowska 85, 61-614 Poznan, Poland\\
$^5$ Institute of Molecular Physics, Polish Academy of Sciences, ul. M. Smoluchowskiego 17, 60-179 Pozna\'{n}, Poland}

\begin{abstract}
In the absence of an external magnetic field and a spin-polarized charge current, an antiferromagnetic system supports two  degenerate magnon modes.  An applied thermal bias activates the magnetic  dynamics, leading to a magnon flow from the hot to the cold edge (magnonic spin Seebeck current). Both degenerate bands contribute  to the magnon current but the  orientations of the magnetic moments underlying  the magnons are  opposite in different bands. Therefore, while the magnon current is nonzero, the net spin current is zero. To obtain a nonzero net spin  current, one needs to apply either a magnetic field or a spin-polarized charge current that lifts the bands' degeneracy.  Here,  attaching a thermal contact to  one edge of a helical nanowire, we study three different magnonic spin currents: (i) the exchange, and (ii) Dzyaloshinskii–Moriya spin currents flowing along the helical nanowire,  and (iii) magnonic spin current pumped into the adjacent normal metal layer. We find that the combination of  Dzyaloshinskii–Moriya interaction and external  magnetic field  enhances substantially the spin current compared to the current generated solely through a magnetic field. Due to  nonreciprocal magnons and magnon dichroism effect, the Dzyaloshinskii–Moriya and the exchange spin currents show left-right propagation asymmetry, with  $20\%$ of  current rectification. The spin pumping current shows a slight asymmetry only in the case of a strong Dzyaloshinskii–Moriya interaction. The observed effects are explained in terms of the magnon dispersion relations and the magnon Doppler effect.
\end{abstract}
\date{\today}
\maketitle
\section{Introduction}

Since the discovery of spin Seebeck effect (SSE),  ferromagnetic insulators  have been in the focus of spin-caloritronic research \cite{uchida2008observation,uchida2010spin,
uchida2010observation,xiao2010theory}.  Relatively less attention were devoted to antiferromagnetic (AFM)  spin-caloritronic research, even though
AFM materials  possess  favorable properties  for spintronic applications (such as  a switching frequency in the  terahertz regime and the absence of stray fields) \cite{qaiumzadeh2017spin,cheng2018giant,hals2011phenomenology,
	jungwirth2016antiferromagnetic,khymyn2016transformation,gomonay2018spin,gomonay2010spin,
	rezende2016diffusive,PhysRevB.95.014434,takashima2018nonreciprocal,seki2015thermal,wu2016antiferromagnetic, lin2016enhancement,lin2016enhancement,gray2019spin,li2020spin,sonin2010spin,rezende2016theory,tveten2013staggered}.
 Under certain conditions, an  AFM ordering  can  be thought of as being composed of two-sublattices with  opposite  (up and down) ferromagnetic  magnetic order. Even though the magnonic current is finite, the net spin current is zero in this case due to the mutual compensation of the spin currents of the two sublattices. To achieve a finite net spin current  one needs to impose a certain asymmetry between the two   sublattices. Some imbalance can be brought about  by subjecting the sample to an external magnetic filed and/or  a spin-polarized electric current.  The spin current torque or/and the magnetic field lift the spin-degeneracy of the  magnon gas resulting in a finite SSE current in AFM insulators. For various aspects and discussions of the magnonic SSE we refer to the  literature~\cite{koopmans2010m,battiato2010superdiffusive,
chotorlishvili2013fokker,chotorlishvili2019influence,xiao2010theory,
seifert2018femtosecond,adachi2013theory}.

In the present work we propose a method for the generation of magnonic spin current in  helical antiferromagnetic systems. Our approach exploits the system's intrinsic properties, and therefore a technical realization may serve as a low-energy cost and environmentally friendly possibility, in addition to solutions based on spin-polarized charge current. The key point in our approach is the Doppler effect in the dispersion relation for the left-right propagating magnons. The splitting of the magnonic modes can be described by an enhanced effective temperature bias that facilitates  the SSE.
\begin{figure}[htbp]
	\includegraphics[width=0.40\textwidth]{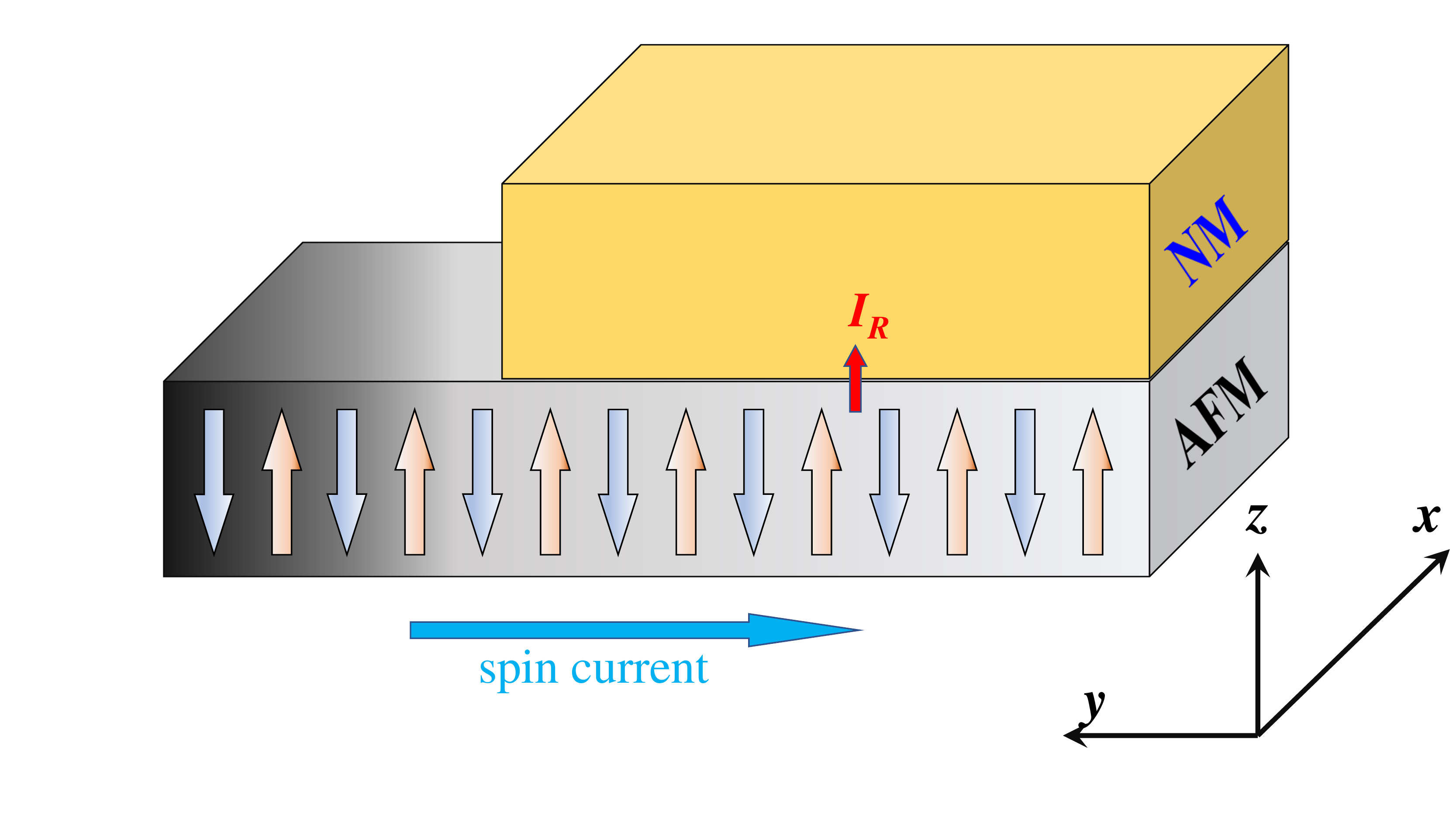}
	\caption{\label{model}  Schematics of the collinear antiferromagnet (AFM) with an attached normal metal (NM). The thermal contact with the temperature $ T_H $  is attached to the left edge of the AFM.  The temperature gradient generates a spin current flowing in the AFM along the $ y $ axis. We specify the exchange $ I_{ex}^z $, the DM $ I_{DM}^z $, and the total  $ I^z $ spin currents.  The thermally activated magnetization dynamics pumps the spin current $I_R$ into the attached NM with the temperature $T_N=0$.  The spin pumping current $ I_R $ is spatially nonuniform since the temperature profile and the magnetization dynamics along the nanowire are not uniform.}
\end{figure}

An asymmetry between the left and right propagating spin waves (magnons) may stem from an antisymmetric exchange interaction, known  as the Dzyaloshinskii–Moriya (DM) coupling. When this interaction is  stronger than a certain threshold value, it may also lead to a helical spin orientation in the corresponding ground state configuration. Below this threshold  of the DM interaction strength, the system is in a collinear ground state. The impact of  DM coupling on  the magnetic ground state and on magnetic excitations (spin waves) in the case of ferromagnets was investigated extensively in  recent years~\cite{tiablikov2013methods,moon2013spin,stagraczynski2017many}. Less attention was paid to DM coupling  in antiferromagnets. But similarly as in ferromagnets, DM interaction in antiferromagnets may lead to a noncollinear ground state, as well. Apart from its influence on the ground state, DM interaction  leads to an asymmetry between the left and right moving magnons.

In the present work, we explore the spin pumping current as well as the spin current flowing in the system that is shown schematically in
Fig.\ref{model}. A hot bath is attached to  the left edge of the AFM insulator, while the right edge temperature is kept at  zero. The thermal bias leads to a magnon
current flowing from the left to the right edge, along the $-\textbf{y}$ direction. If the symmetry between the two AFM modes is broken, this magnon current results in a finite longitudinal spin Seebeck current. In turn, the spin current pumped into the normal metal is actually  the transverse spin current -- flowing along the $\textbf{z}$ axis.

The paper is organized as follows. In section  \textbf{II} we analyze the magnon dispersion relations and the magnon properties in  specific systems with DM interaction. In section \textbf{III} we develop a Fokker-Planck formalism in the macrospin formulation for the effective asymmetric AFM exchange modes. In section \textbf{IV} we present results of  numerical  simulations and study  rectification effects for the exchange spin current in AFM chains. Summary and concluding remarks are in \textbf{V}.

\section{Properties of spin waves}

The broken left-right symmetry in the magnon dispersion relations leads to the asymmetric SSE and rectification of the exchange magnonic spin current. This effect can be interpreted in terms of the magnon dispersion relations.
Therefore, we  consider now properties of spin waves in antiferromagnets with DM interaction, and restrict ourselves to  specific low-dimensional systems, which can be considered as building blocks of two dimensional antiferromagnets.

\subsection{Two ferromagnetic chains coupled antiferromagnetically}

At first, we consider spin waves in two ferromagnetic chains which are exchange-coupled antiferromagnetically and both host DM interaction, see Fig.2. We also include the Zeeman energy due to an external magnetic field  and the contributions due to the magnetic anisotropy with the easy axis being  perpendicular to the chains (same in both chains).
The Hamiltonian describing such a system can be written in  the  form:
\begin{eqnarray}
\label{H_1}
\hat{H} = - \frac{1}{2} J\sum_{\alpha = 1,2} \sum_{<ij>}  \hat{\mathbf{S}}_{\alpha\, i} \cdot \hat{\mathbf{S}}_{\alpha\, j}
- J_{ic} \sum_{i} \hat{\mathbf{S}}_{1i} \cdot \hat{\mathbf{S}}_{2i}\nonumber\\
-  \frac{1}{2} \sum_{\alpha = 1,2} \sum_{<ij>} \mathbf{D}^{\alpha}_{ij} \cdot \left(\hat{\mathbf{S}}_{\alpha \, i} \times \hat{\mathbf{S}}_{\alpha\, j} \right)\nonumber \\
- K \sum_{\alpha =1,2} \sum_{i} (\hat{S}_{\alpha\, i}^{z})^2 - H_z \sum_{\alpha =1,2} \sum_{i}  \hat{S}_{\alpha\,i}^{z}\;, \hspace{0.5cm}
\end{eqnarray}
 where the index $\alpha=1$, or $2$ distinguishes the two chains, $J>0$ is the ferromagnetic exchange coupling parameter  between nearest neighbours
within the chains, and $J_{ic}<0$ is the strength  of antiferromagnetic inter-chain coupling.  $K$ quantifies  the anisotropy energy  (assumed to be the same in both chains), $\mathbf{D}^{\alpha}_{ij}$ is the DM vector  in the $\alpha$-th chain, while $H_z$ is the external  magnetic field measured in energy units ($H_z=\gamma\hbar{\tilde{H}}_z$, where $\gamma$ is the gyromagnetic ratio and ${\tilde{H}}_z$ is the unnormalized magnetic field). The summation in the intra-chain exchange and DM terms is over the  nearest neighbors and the factors ($1/2$) are introduced to cancel double counting of interacting spin pairs.

\begin{figure}[h]
    \centering
    \includegraphics[width=0.9\columnwidth]{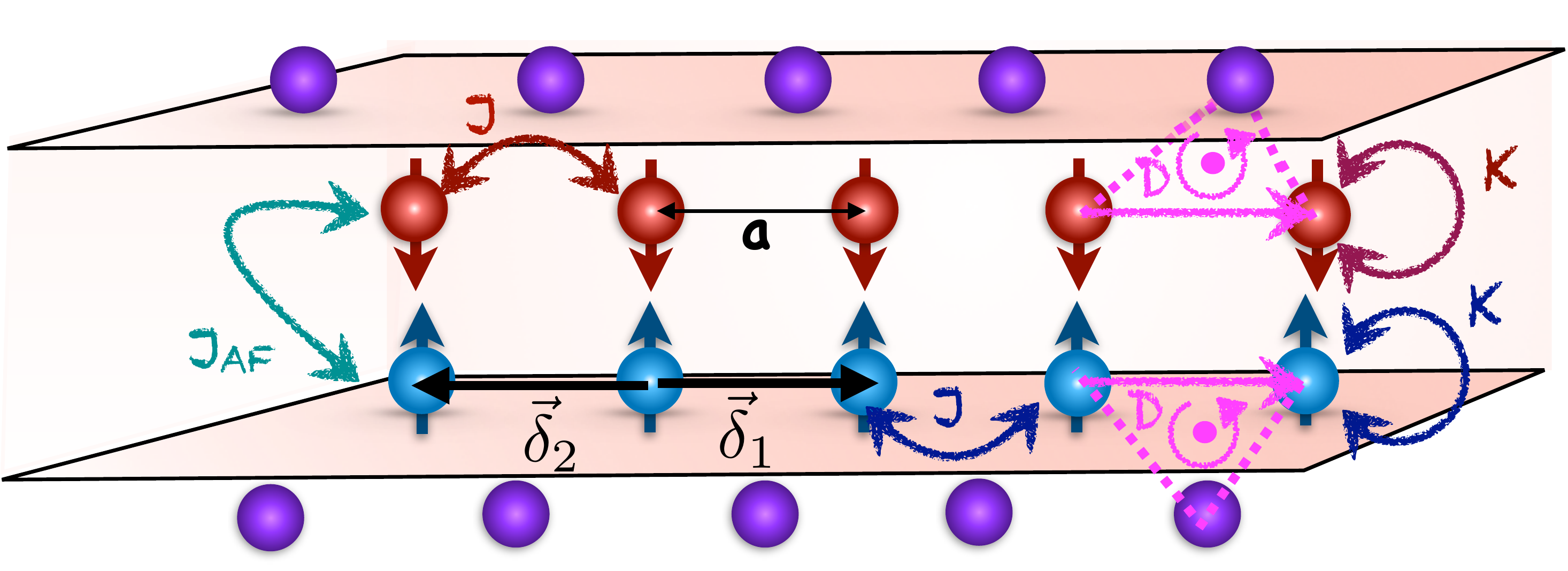}
    \caption{Schematics of two ferromagnetic chains coupled antiferromagnetically by exchange  interactions. Both chains host DM interactions. }
    \label{fig:1}
\end{figure}

For simplicity, we limit the following considerations to the collinear spin ground state, which is stabilized by the easy-axis magnetic anisotropy (assumed to be along the axis $z$ normal to the chains) as well as by the antiferromagnetic inter-chain exchange coupling. Such a stable collinear state appears when the DM interaction and also the external field (applied along the easy axis) are small compared to the magnetic anisotropy, and are smaller than  certain threshold values. In a more general case, however,  a noncollinear ground state configuration is stable. Generally, the DM parameters in both chains can be different, but at first we restrict the consideration to the case  $\mathbf{D}_{ij}^{\alpha =2} =  \mathbf{D}_{ij}^{\alpha =1} = \mathbf{D}_{ij}$ and $\mathbf{D}_{ij} = \mathcal{D}\xi_{ij} \hat{\mathbf{z}}$, with $\xi_{ij}$ being defined as  $\xi_{ij}= 1$ for $j=i+1$ and $\xi_{ij}= -1$ for $j=i-1$. Upon a Holstein transformation, followed by a Fourier transform,
we arrive at the following bi-linear Hamiltonian:
\begin{eqnarray}
\label{Hkappa1}
\hat{H} = \sum_{\mathbf{k}} \left[[\mathcal{F}(\mathbf{k})-\mathcal{G}(\mathbf{k})] (\hat{a}_{\mathbf{k}}^{+}\hat{a}_{\mathbf{k}} + \hat{b}_{\mathbf{k}}^{+}\hat{b}_{\mathbf{k}}) \right.\nonumber \\
\left. +  H_z (\hat{a}_{\mathbf{k}}^{+}\hat{a}_{\mathbf{k}} - \hat{b}_{\mathbf{k}}^{+}\hat{b}_{\mathbf{k}})
+ \mathcal{C} (\hat{a}_{\mathbf{k}}^{+} \hat{b}_{-\mathbf{k}}^{+} + \hat{a}_{\mathbf{k}} \hat{b}_{-\mathbf{k}}) \right].
\end{eqnarray}
A constant term is  suppressed, $\mathbf{k}$ is the one-dimensional wavevector, and   the following definitions are introduced
\begin{align}
\label{F}
\mathcal{F}(\mathbf{k}) &= -  J S \sum_{\bdelta}\left[\cos(\mathbf{k}\cdot\bdelta) - 1\right] - J_{ic}S + A S , \nonumber \\
& \mathcal{G}(\mathbf{k})=    S \mathcal{D} \sum_{\bdelta} \xi_{\bdelta} \sin(\mathbf{k}\cdot\bdelta), \nonumber  \\
& \mathcal{C} = |J_{ic}| S.
\end{align}
Here,  $\xi_{\bdelta}$ is defined as $\xi_{\bdelta}=1$ for the vector $\bdelta$  connecting the spins at sites $i$ and $i+1$, while
   $\xi_{\bdelta}=-1$ stands for $\bdelta$ connecting the spins  at sites $i$ and $i-1$, and  $A=2K$ is the anisotropy field (in energy units).

Performing  the Bogolyubov transformation,
\begin{eqnarray}
\hspace*{-0.5cm}\left\{\begin{array}{l}
 \hat{a}_{\mathbf{k}} = u_{\mathbf{k}} \hat{c}_{\mathbf{k}} + \nu_{\mathbf{k}} \hat{d}_{-\mathbf{k}}^{+}    \\
 \\
\hat{a}_{\mathbf{k}}^{+} = u_{\mathbf{k}} \hat{c}_{\mathbf{k}}^{+} + \nu_{\mathbf{k}} \hat{d}_{-\mathbf{k}}
\end{array}\right. \quad  \quad \left\{\begin{array}{ll}
\hat{b}_{\mathbf{k}} = u_{\mathbf{k}} \hat{d}_{\mathbf{k}} + \nu_{\mathbf{k}} \hat{c}_{-\mathbf{k}}^{+}\\
\\
\hat{b}_{\mathbf{k}}^{+} = u_{\mathbf{k}} \hat{d}_{\mathbf{k}}^{+} + \nu_{\mathbf{k}} \hat{c}_{-\mathbf{k}}
\end{array}\right. .
\end{eqnarray}
The coefficients  $u_{\mathbf{k}}$ and $\nu_{\mathbf{k}}$   obey the relation
$ u_{\mathbf{k}}^{2} - \nu_{\mathbf{k}}^{2} = 1$,
while $\hat{c}_{\mathbf{k}}$ ($\hat{c}_{\mathbf{k}}^{+}$) and $\hat{d}_{\mathbf{k}}$ ($\hat{d}_{\mathbf{k}}^{+}$) are the new bosonic annihilation (creation) operators. The  Hamiltonian reads
\begin{align}
\label{Hcd}
\hat{H} = \sum_{\mathbf{k}}\left[ \mathcal{W}(\mathbf{k})  (u^{2} + \nu^{2}) +   H_z + 2 u \nu \mathcal{C}\right] \hat{c}_{\mathbf{k}}^{+} \hat{c}_{\mathbf{k}}\nonumber\\
+  \sum_{\mathbf{k}}\left[ [\mathcal{W}(\mathbf{k})  (u^{2} + \nu^{2}) -   H_z + 2 u \nu \mathcal{C}\right] \hat{d}_{\mathbf{k}}^{+} \hat{d}_{\mathbf{k}}\nonumber\\
+ \sum_{\mathbf{k}} \left[2 \mathcal{W}(\mathbf{k}) u \nu + \mathcal{C} (u^{2} + \nu^{2}) \right] (\hat{c}_{\mathbf{k}}^{+} \hat{d}_{-\mathbf{k}}^{+} + \hat{c}_{\mathbf{k}} \hat{d}_{-\mathbf{k}}),
\end{align}
where $\mathcal{W}(\mathbf{k}) =\mathcal{F}(\mathbf{k})-\mathcal{G}(\mathbf{k})$.
The
Hamiltonian (\ref{Hcd}) becomes diagonal when the following condition is fulfilled
\begin{eqnarray}
2 \mathcal{W}(\mathbf{k}) u \nu + \mathcal{C} (u^{2} + \nu^{2}) = 0,
\end{eqnarray}
and the Hamiltonian can be written as (see Appendix)
\begin{equation}
    \hat{H} =  \sum_{\mathbf{k}} \left(\varepsilon_{\mathbf{k}1}\, \hat{c}_{\mathbf{k}}^{+}\hat{c}_{\mathbf{k}} + \varepsilon_{\mathbf{k}2}\, \hat{d}_{\mathbf{k}}^{+}\hat{d}_{\mathbf{k}} \right),
\end{equation}
with the eigenmode dispersion 
\begin{align}
    \varepsilon_{\mathbf{k}1,2} &= \sqrt{\mathcal{W}^{2}(\mathbf{k}) - \mathcal{C}^{2}} \pm  H_z.
\end{align}
Considering Eqs. (3) and (4) and assuming that the contribution due to DM interaction is small, one finds
\begin{eqnarray}
    \varepsilon_{\mathbf{k}1,2} = \sqrt{\varepsilon_{exA}(\varepsilon_{exA}+2\varepsilon_{AF})}\nonumber \\
    - \frac{\varepsilon_{exA}+\varepsilon_{AF}}{\sqrt{\varepsilon_{exA}(\varepsilon_{exA}+2\varepsilon_{AF})}}\varepsilon_{DM}\pm H_z ,
\end{eqnarray}
with
\begin{eqnarray}
    \varepsilon_{exA} = 2 J S \left(1 - \cos{(k_{x} a)}\right) + AS, \nonumber \\
     \varepsilon_{DM} = 2  \mathcal{D} S \sin{(k_{x}a)}, \nonumber \\
     \varepsilon_{AF} =\mathcal{C}=|J_{ic}|S.
\end{eqnarray}
When the anisotropy term is dominant,  the eigen-energies in the  collinear regime can be written as
\begin{equation}
    \varepsilon_{\mathbf{k}1,2}
    \approx  \varepsilon_{exA}  + \varepsilon_{AF} - \varepsilon_{DM} \pm  H_z.
\end{equation}
As follows from the above, the external field splits the two magnon branches. The DM coupling leads to an asymmetry between the left and the right propagating magnons (Doppler effect). Since the DM parameter was assumed to be of the same strength  for both chains, the Doppler shift, described by $\mathcal{D} S \sin{(k_{x}a)}$ is the same for both modes.

When the DM parameters have different signs  in  two chains, i.e., 
$\mathbf{D}_{ij}^{\alpha =2} = -\mathbf{D}_{ij}^{\alpha =1} = -\mathbf{D}_{ij}$, we find upon similar considerations 
\begin{equation}\label{the DM parameters are opposite}
    \varepsilon_{\mathbf{k}1,2} = \sqrt{\varepsilon_{exA}(\varepsilon_{exA}+2\varepsilon_{AF})}\mp \varepsilon_{DM}\pm H_z.
\end{equation}
For a strong anisotropy contribution 
\begin{equation}\label{for a strong anisotropy}
    \varepsilon_{\mathbf{k}1,2}
    \approx  \varepsilon_{exA}  + \varepsilon_{AF} \mp \varepsilon_{DM} \pm  H_z
\end{equation}
 applies. As follows from Eq.~(\ref{the DM parameters are opposite}) and Eq.(\ref{for a strong anisotropy}), the DM term leads  not only to a magnon Doppler effect, but also lifts the degeneracy of the spin waves. The general case, $\mathbf{D}_{ij}^{\alpha =2} = \kappa \mathbf{D}_{ij}^{\alpha =1}$, is briefly presented in the Appendix.

\subsection{Single antiferromagnetic chain}
We consider  a single antiferromagnetic chain of localized spins using the following Hamiltonian, restricted to the nearest-neighbor interactions:
\begin{eqnarray}\label{Free energy}
&&\hat{H}_{s}=\frac{1}{2}\sum_{\langle n, m \rangle} I \hat{S}_n \cdot \hat{S}_{m} - \frac{1}{2}\sum_{\langle n, m \rangle} \vec{D}_{nm}\cdot\left(\hat{S}_{n} \times \hat{S}_{m} \right)  \nonumber\\ 
&& - K \sum_n ({\hat{S}}^z_n)^2 - \sum_n {\bf H} \cdot \hat{S}_{n}.
\end{eqnarray}
The exchange interaction constant $ I > 0 $ is antiferromagnetic,  while the DM interaction  can be written as, $ \vec{D}_{nm} =D\xi_{nm}\hat{z}$, where $D$ is the  coupling strength, $\hat{z}$ stands for a unit vector along the axis $z$ (easy axis),  and  $ \xi_{nm}=\pm1 $  for $m=n\pm 1$. 
The last two terms have the same meaning as in Eq.(1). 

The spin dynamics is governed by the Landau-Lifshitz-Gilbert (LLG) equation,
\begin{eqnarray}\label{OLLG}
&& \hbar \frac{\partial \hat{S}_n}{\partial t} = \hat{S}_n \times \frac{\delta \hat{H}_{s}}{\delta \hat{S}_n} + \frac{\hbar \alpha}{S} \hat{S}_n \times \frac{\partial \hat{S}_n}{\partial t},
\end{eqnarray}
where $\alpha$ is the Gilbert damping constant. We performed numerical calculations based on Eq. (\ref{OLLG}) for a one
dimensional (1D) chain consisting of 100 sites, using  the following parameters:  The exchange constant $ I = 5.5 \times 10^{-21} $ J,   $ S = 0.2 $,  the anisotropy constant $ K = 4.4 \times 10^{-23} $ J, and the damping constant $ \alpha = 0.001 $. These parameters are relevant for uniaxial AFM materials, like for instance  MnF$_2$.  We note, that relying on earlier studies reported in  Ref. \cite{PhysRevB.102.104436}, the model under considerations is appropriate to describe the experimental observations induced by the combined action of DMI and magnetic field.
\begin{figure}[htbp]
	\includegraphics[width=0.48\textwidth]{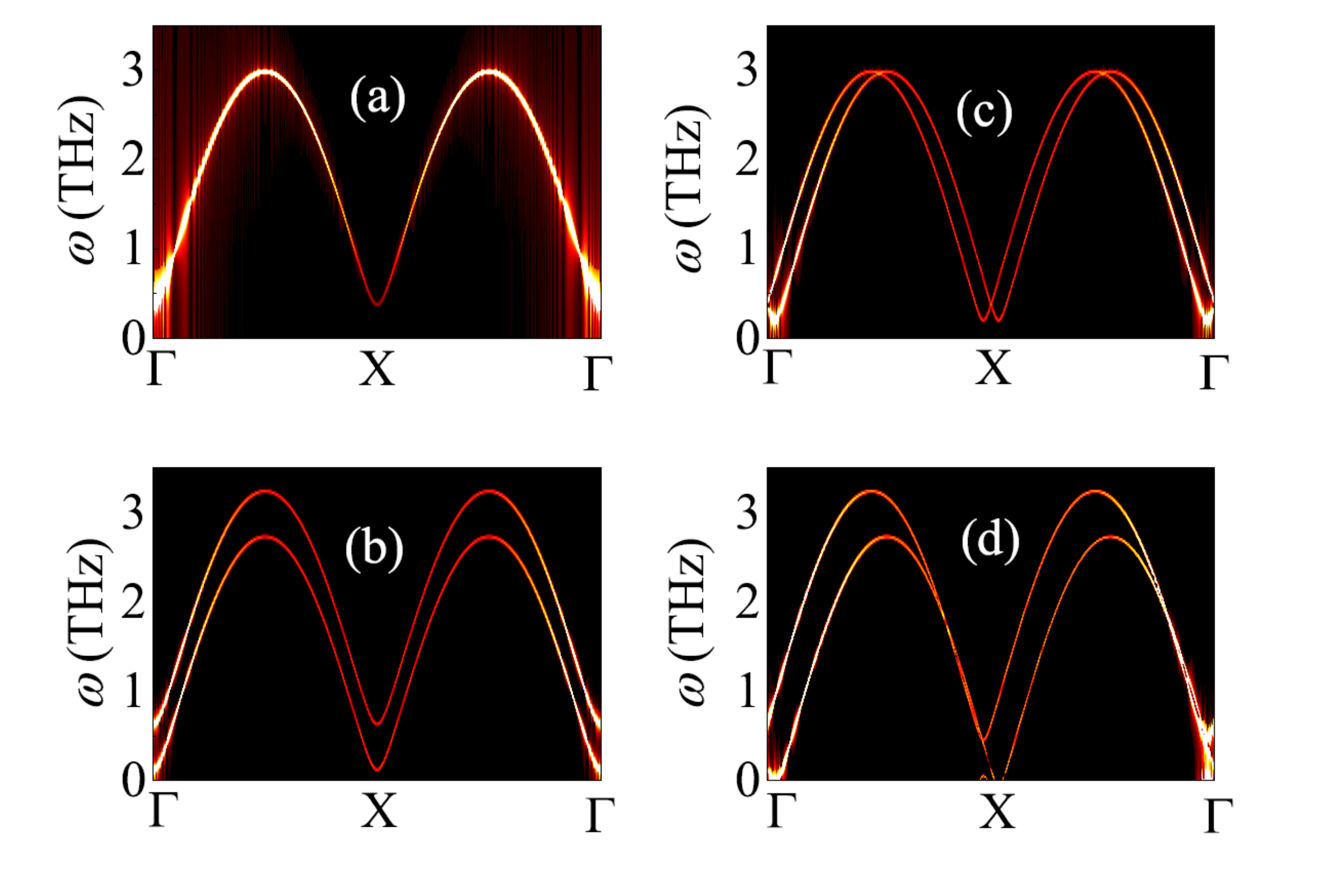}
	\caption{\label{dispersion}  The magnon dispersions in the collinear AFM spin chain when: (a) $ H_z = 0 $ and $ D = 0 $, (b) $ H_z = 1.7 \times 10^{-22} $ J and $ D = 0 $, (c) $ H_z = 0 $  and $ D = 5.9 \times 10^{-22} $ J, and (d) $ H_z = 1.7 \times 10^{-22} $ J and $ D = 5.9 \times 10^{-22} $ J. }
\end{figure}

From numerical simulations  we obtained the magnon dispersion relations in the helical antiferromagnetic chain, presented in Fig. \ref{dispersion}. To calculate these dispersion relations, we assume a pulse of  periodic magnetic field applied along the $ \textbf{x} $ direction, $ H_x(t) = h_0 \sin(\omega t) / (\omega t) $, with the amplitude $ h_0 = 6.4 \times 10^{-22} $ J and frequency $ \omega = 4 $ THz. The pulse of magnetic field is applied locally to the center of the sample.  During the  numerical calculations, for each sublattice we extract the value of $ S_x$  during the period of 50 ps and a time step of 5 fs. The dispersion relation is obtained through a two-dimensional fast Fourier transform, $ S_x(k, \omega) =  F_2[S_x(n, t)] $.

In the absence of a magnetic field and DM interactions, the two magnon modes of frequency $ \omega_+ $ and $ \omega_- $ are degenerate, see Fig. \ref{dispersion}(a). The lowest energy in the spectrum corresponds to the wavevector $ k = 0 $, and also to the boundaries of the Brillouin zone, $ k = \pm \pi/a $, where $ a $ is the lattice parameter (distance between the neighboring sites in the chain).  This energy is determined by the magnetic anisotropy. The obtained dispersion relations are in agreement with the results of  Ref. \cite{PhysRevB.102.104436}.

The external magnetic field along the easy-axis ($ \textbf{z} $ axis) lifts the degeneracy, see Fig. \ref{dispersion}(b). One of the modes is shifted upwards, while the other one is shifted downwards. For the assumed magnetic field  $ H_z = 1.7 \times 10^{-22} $ J, the collinear ground state is stable, as the field obeys the stability condition  $ H_z < 2S \sqrt{I K} = 1.8 \times 10^{-22} $ J. The DM interaction lifts the degeneracy as well, as demonstrated in Fig. \ref{dispersion}(c).  The assumed magnitude  of the DM interaction constant, $ D = 5.9 \times 10^{-22} $ J,  corresponds to the stable collinear ground state  ($ D < \sqrt{(4I + K) K} = 9.8 \times 10^{-22} $ J).  The two magnon modes are horizontally shifted in opposite directions, so that the lowest frequencies are moved away from the center and the boundaries of the Brillouin zone. 

In the absence of a  magnetic field, the magnon spectrum preserves the left-right mirror symmetry with respect to  the inversion operation, Fig. \ref{dispersion}(c). This can be accounted for  as follows. The inversion operation changes the sign of the DM term. The positive sign of the constant $ D $ for the upper layer with positive equilibrium magnetization is equivalent to the $-D $ and negative equilibrium magnetization of the bottom layer. Thus, the inversion operation mimics a switching of the layers, and the left-right symmetry holds as long as  the  layers are equivalent.

The applied external magnetic field $ H_z = 1.7 \times 10^{-22} $ J, together with the DM interaction $ D = 5.9 \times 10^{-22} $ J,   breaks the
mirror symmetry,  see Fig. \ref{dispersion}(d). The gap for one of the bands is reduced, while that  for the other band is increased. Due to the breaking of the left-right symmetry and the deformation of the band structures, the contributions of the two magnon modes to the SSE are different.  In what follows, we explore this asymmetry in more details.

\begin{figure}[htbp]
	\includegraphics[width=0.48\textwidth]{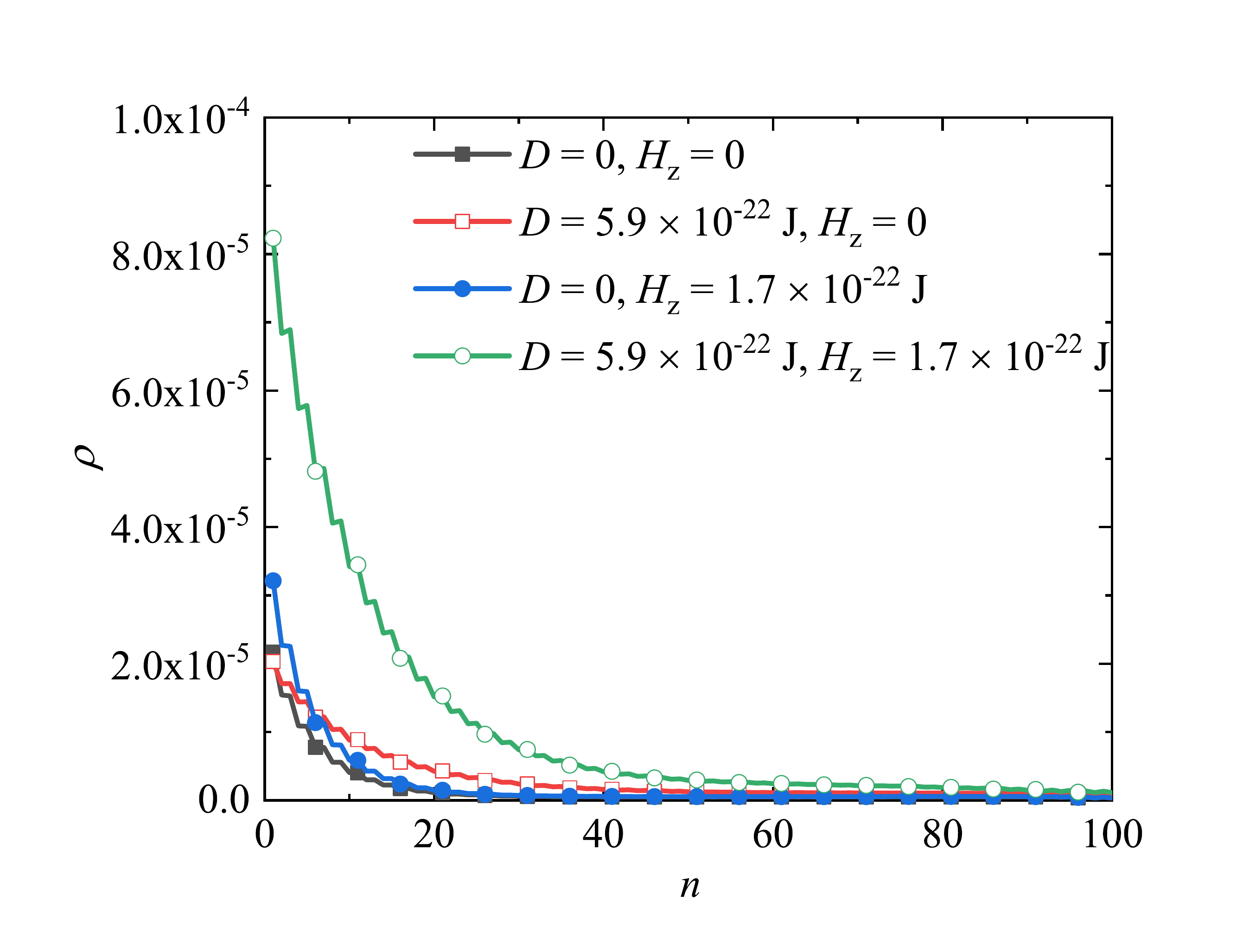}
	\caption{\label{profile-rou} Spatial profile of the thermal magnon density $ \rho = m_x^2 + m_y^2 $. Here, $ m_x $ and $ m_y $ are the $ x $ and $ y $ components of the local magnetization unit vector. The temperature  at the left edge of the AFM chain is $ T = 30 $ K and $ T = 0 $ at the right edge. The temperature profile (magnon density) formed in the AFM is not spatially uniform and decays from the left edge on a length scale that is larger than the magnon propagation length.}
\end{figure}

 We consider the case where the temperature drops from $ T = 30 $ K at the left edge to  $ T = 0 $ at right edge of the AFM. The thermally excited magnons propagate then from the left edge to the right one. Let us address the issue of  how the magnetic field and DM interactions influence  the magnon density.  The density of  thermally excited magnons is determined through the transversal components, $ \rho = m_x^2 + m_y^2 $, of the local magnetization unit vector (the $ \textbf{z} $ axis is along the direction of equilibrium magnetization). As follows from Fig. \ref{profile-rou}, in the absence of a magnetic field,  the DM interaction  enhances slightly the magnon density $ \rho $. However, when a magnetic field is applied, the effect becomes relatively large.  The DM term and the magnetic field generate together a larger gradient of the magnon density. Since the temperature is nonzero only at the left edge, far away from this edge -- at a distance larger than the magnon propagation length -- the magnon density approaches  zero.  Thus, to obtain a sizable  pumping current $ I_R $ in AFM, one needs both the DM interaction and the magnetic field.

\section{Fokker-Planck formulation}

 To calculate the  AFM magnonic current    we  utilize the Fokker-Planck equation. To this end, we consider an effective model based on 
two macrospins which are coupled through an exchange mean field. The two macrospins represent the two spin sublattices. One may consider the macrospins as a homogeneous part of the double chain studied in Sec.2.1.    
In a simplified description, the precession states of the macrospins can be approximated by the spin wave modes given by Eq.(13), where for simplicity we  neglect the anisotropy term. In this section we write frequencies of these modes,
$\hbar \omega_\pm(\textbf{k},D) \equiv \varepsilon_{\mathbf{k}1,2}$, in the form
$\omega_\pm(\textbf{k},D)=\omega_0(\textbf{k})\pm\omega_0\mp\omega_D(\textbf{k})$, where  $\omega_D(\textbf{k})=2S{\cal{D}}\sin(k_xa)$, $\omega_0(\textbf{k})=[2JS(1-\cos(k_xa)+ \varepsilon_{AF})/\hbar$, and $\omega_0=\gamma {\tilde{H}}_z$.
We also assume that the effect of  magnetic field is larger than that of the DM interaction, but smaller than the contribution  due to exchange interaction,  $\omega_0(\textbf{k})>\omega_0 >\omega_D$.

 Considering the magnonic current associated with these two modes, we note  that 
the transverse magnonic modes contributing to the magnonic current show nontrivial damping properties. In particular, the damping depends on the dispersion relations
and the wave vector \cite{kajiwara2010transmission} as follows: $\textbf{M}_\pm^\bot\left(\textbf{r},t\right)\propto\exp\left(i\textbf{k}\textbf{r}+i\omega_{\pm}(\textbf{k})t\right)\exp\left(-\alpha\omega_{\pm}(\textbf{k})t\right)$, where $\alpha$ is the Gilbert damping constant. To find  the magnetization associated with a particular mode, we use the LLG equation: 
\begin{equation}
\displaystyle \frac{\partial \vec{m}_{\pm}}{\partial t} = - \gamma \vec{m} \times \bigg(\tilde{\vec{H}}^{\mathrm{\rm eff}}_{\pm} + \tilde{\vec{h}}_l\bigg)+ \alpha \vec{m}_{\pm} \times \frac{\partial \vec{m}_{\pm}}{\partial t}.
\label{LLG}
\end{equation}
Here, the  unit vectors of the magnetization are defined as follows: $\textbf{m}_{\pm}=\textbf{M}_{\pm}/M_s$, and $M_s$ is the saturation magnetization. We recall that we utilize  the approximation of  two macrospins with  asymmetric left-right ($-,+$) effective modes. The meaning of Eq.(\ref{LLG}) is that each magnetic moment experiences the local effective magnetic field \cite{gomonay2010spin}. The simplest case concerns the system in the vicinity of equilibrium, where all the possible motions of the AFM vector could be represented in terms of two noninteracting normal modes \cite{gomonay2013peculiarities}. Analysis  of the intrinsic cutoff frequencies of the time-resolved SSE shows that mainly sub-thermal small wavelength magnons contribute to the SSE \cite{etesami2015spectral}.
On the other hand, the magnetic ordering is antiferromagnetic and the effective field acting on the magnetization of each (left-right) mode can be presented in the form 
$\gamma{\tilde{\vec{H}}}^{\mathrm{\rm eff}}_{1,2}=(0,0,\omega_{\pm}(\textbf{k}, D))$.
The temperature in the LLG equation enters through the correlation function of the thermal random magnetic field $\tilde{\vec{h}}_l$,  $ \langle {\tilde{h}}_{l,p}(t,\vec{r}) {\tilde{h}}_{l,q}(t',\vec{r}') \rangle = \frac{2 k_B T \alpha}{\gamma \mu_0 M_s V} \delta_{pq} \delta(\vec{r}-\vec{r}') \delta(t-t') $, where $ p,q = x,y,z $, while $ k_B $ is the Boltzmann constant and $ V $ is the volume of the unit cell.

For the sake of simplicity, we assume here that the dominant fields 
acting on the magnetic moments are the external magnetic field and the exchange field. Therefore, the effective field is aligned along $\textbf{z}$ axis. The DMI field strength is weak and the corresponding contribution is  treated perturbatively. The weak anisotropy field is neglected throughout this section. The spin currents generated by the individual modes read \cite{chotorlishvili2013fokker,chotorlishvili2019influence}
\begin{equation}\label{spincurrent}
\langle\vec{I}\rangle_{\pm}=\frac{M_{s}V}{\gamma}\alpha'\langle\vec{m}_{\pm}\times\dot{\vec{m}}_{\pm}\rangle,
\end{equation}
where $\alpha^\prime =\gamma\hbar g_{r}/4\pi M_{s}V$ with the real part of the
mixing spin conductance constant $g_{r}$.
The total spin current \cite{PhysRevB.92.134424} reads:
\begin{equation}\label{spincurrent2}
\langle\vec{I}\rangle_{R,\textbf{k}}=\frac{M_{s}V}{\gamma}\alpha'\left\lbrace \langle\vec{m}_{-}\times\dot{\vec{m}}_{-}\rangle-\langle\vec{m}_{+}\times\dot{\vec{m}}_{+}\rangle\right\rbrace.
\end{equation}

The currents associated with the individual modes have been calculated from the FP equation
\cite{chotorlishvili2013fokker,chotorlishvili2019influence}. We omit technical details and
present the final result. The partial spin current generated by  magnons with the wave vector $\textbf{k}$ reads:
\begin{widetext}
\begin{eqnarray}\label{total spin current}
&& \langle I\rangle_{R,\textbf{k}}=\frac{M_sV\alpha'}{\gamma}\left\lbrace\omega_{+}(\textbf{k},D)\langle(m_+^z)^2\rangle- \omega_{-}(\textbf{k},D)\langle(m_-^z)^2\rangle\right\rbrace,\nonumber\\
&& \langle(m_{\pm}^z)^2\rangle=1-\frac{2}{\beta\hbar\omega_{\pm}(\textbf{k},D)}L\left(\beta
\hbar\omega_{\pm}(\textbf{k},D)\right),\nonumber\\
\end{eqnarray}
\end{widetext}
where 
the inverse temperature is given by
\begin{eqnarray}\label{total spin current2}
&& \beta=\frac{2\alpha\left(1+\alpha^2\right)}{\sigma^2\hbar},\nonumber\\
&&\sigma^2=\frac{2\alpha\gamma k_{B}T_{m}}{M_{s}V},
\end{eqnarray}
while $T_{m}$ is the magnon temperature, and $L\left(\beta\omega_{\pm}(\textbf{k})\right)$ is the Langevin function.\\

The obtained result can be interpreted in terms of the effective inverse helical temperatures $\beta_{\pm}^{\rm eff}=\beta\left(1\pm \frac{Dak}{\omega_0(\textbf{k})\pm\omega_0}\right)$.
For convenience, we rewrite Eq.(\ref{total spin current}) in the following form:
\begin{widetext}
\begin{eqnarray}\label{total spin current3}
&& \langle I\rangle_{R,\textbf{k}}=\frac{M_sV\alpha'}{\gamma}\left\lbrace\omega_0(\textbf{k})
\langle(m_+^z)^2\rangle-
\langle(m_-^z)^2\rangle)+(\omega_0+akD)(\langle(m_+^z)^2\rangle+\langle(m_-^z)^2\rangle)\right\rbrace,\nonumber\\
&& \langle(m_+^z)^2\rangle-\langle(m_-^z)^2\rangle=\frac{4(\omega_0+akD)}{\beta\hbar\omega_0^2(\textbf{k})}\left(L\left(\beta\hbar\omega_0(\textbf{k})\right)-L'\left(\beta\hbar\omega_0(\textbf{k})\right)\right),\nonumber\\
&&\langle(m_+^z)^2\rangle+\langle(m_-^z)^2\rangle=2+\frac{4(\omega_0+akD)^2}{\beta\hbar\omega_0^3(\textbf{k})}
L'\left(\beta\hbar\omega_0(\textbf{k})\right)-\frac{4}{\beta\hbar\omega_0(\textbf{k})}L\left(\beta\hbar\omega_0(\textbf{k})\right).
\end{eqnarray}
\end{widetext}
In the low-temperature limit $\beta\hbar\omega_0(k)>1$,  Eq.(\ref{total spin current3}) simplifies and the expression for the spin Seebeck current reads 
\begin{widetext}
\begin{eqnarray}\label{final SSELOW}
\langle I\rangle_{R,\textbf{k}}=\frac{M_sV\alpha'}{\gamma}\left\lbrace \frac{4(\omega_0+akD)}{\beta\hbar\omega_0(k)}+2(\omega_0+akD)\left(1-\frac{2}{\beta\hbar\omega_0(k)}\right)\right\rbrace .
\end{eqnarray}
\end{widetext}
In the high temperature limit $\beta\hbar\omega_0(k)<1$
\begin{widetext}
\begin{eqnarray}\label{final SSEHIGH}
\langle I\rangle_{R,\textbf{k}}=\frac{M_sV\alpha'}{\gamma}\left\lbrace \frac{4(\omega_0+akD)}{3\beta\hbar\omega_0(k)}(\beta\hbar\omega_0(k)-1)+(\omega_0+akD)\left(\frac{4(\omega_0+akD)^2}{3\beta\hbar\omega_0^2(k)}-2/3\right)\right\rbrace .
\end{eqnarray}
\end{widetext}
As we see from Eq.(\ref{final SSELOW}) and Eq.(\ref{final SSEHIGH}), the spin Seebeck current vanishes in the absence of an external magnetic field and DM interaction.

\section{Magnonic current in an extended system}

Now we consider magnonic spin  currents in an extended system, i.e., in a  long chain that includes many atomic sites, so the macrospin approximation becomes inappropriate. 
Contrary to the Sec.3, the  current is  spatially non-uniform in the general case  due to the temperature gradient. For the numerical calculations we use the  method developed  in Ref.\cite{PhysRevB.90.014410}.

In what follows, we calculate several magnonic spin currents defined as:\\

\textbf{[i]}. Exchange spin current $I_{ex}^z$ and DM spin current $I_{DM}^z$.  Both currents flow along the nanowire, in the $-\textbf{y}$ direction when the thermal bias is applied (temperature $T=30$K is applied at the left edge of the AFM). In the case of a uniform temperature, both currents are zero.  We note that for our system, a temperature of 30K is well above the regime  where quantum effects might be important \cite{PhysRevB.103.054436}. Therefore, such effects are deemed unimportant here.\\
\textbf{[ii]}. The total spin current is the sum of the exchange and the DM currents, $I^z=I_{ex}^z+I_{DM}^z$. We will show that the exchange, DM, and total current are maximal when both the DM interaction and the  magnetic fields are present.  All three magnonic currents show rectification effects, i.e., the current for $-D $ is larger than that for $ D $. Note, a switching the sign of the  DM constant is equivalent to a left-right inversion.\\
\textbf{[iii]}. The spin pumping current, $I_{R}$, is the current pumped from the AFM to the normal metal, see  Eq.(\ref{spincurrent}). We note that this current is nonzero even if the AFM temperature is uniform. When the AFM temperature is nonuniform, the spatial profile of the spin pumping current is also nonuniform.\vspace{0.2cm}

Following  Ref. \cite{PhysRevB.90.014410}, the expression for the exchange spin-current tensor obeys the equation,
\begin{widetext}
	\begin{equation}
	\displaystyle \vec{\nabla} \textbf{j}_{ex} = -\frac{1}{\hbar} \hat{S}_n \times \frac{\delta \hat{H}_{ex}}{\delta \hat{S}_n},
	\label{exchangecurrent}
	\end{equation}
\end{widetext}
where $ \hat{H}_{ex} = \frac{1}{2}\sum_{\langle n, m \rangle} I \hat{S}_n \cdot \hat{S}_{m} $ is the Hamiltonian for the exchange coupling. In the sublattice model, we consider a discrete version of the gradient operator and the exchange spin-current tensor from the helical antiferromagnetic system  is,
\begin{equation}
I_{n,ex}^\alpha=I_{0,ex}^\alpha-2\frac{J}{\hbar}\sum\limits_{p=1}^n S_p^\beta(S_{p-1}^\gamma+S_{p+1}^\gamma)
\varepsilon_{\alpha\beta\gamma},
\label{desecrate formula}
\end{equation}
where $\varepsilon_{\alpha\beta\gamma}$ is the Levi-Civita antisymmetric tensor, while the Greek indices define the current components. Similarly, for the DM current, using the Hamiltonian $\hat{H}_{DM} = - \sum_{\langle n, m \rangle} \vec{D}_{nm}\cdot\left(\hat{S}_{n} \times \hat{S}_{m} \right)$ and $\vec{\nabla} \textbf{j}_{DM} = -\frac{1}{\hbar} \hat{S}_n \times \frac{\delta \hat{H}_{DM}}{\delta \hat{S}_n}$,  we find
\begin{widetext}
	\begin{equation}
	I_{n,DM}^\alpha=I_{0,DM}^\alpha - 2 \frac{D}{\hbar} \sum\limits_{p=1}^n [S_p^\beta(S_{p+1}^\beta - S_{p-1}^\beta) + S_p^\gamma(S_{p+1}^\gamma - S_{p-1}^\gamma)].
	\label{dmicurrent}
	\end{equation}
\end{widetext}

We consider two situations. The first one is a system with a uniform temperature. The second situation refers to a system with a temperature gradient. 

\subsection{Uniform temperature: spin pumping current}

\begin{figure}[htbp]
	\includegraphics[width=0.48\textwidth]{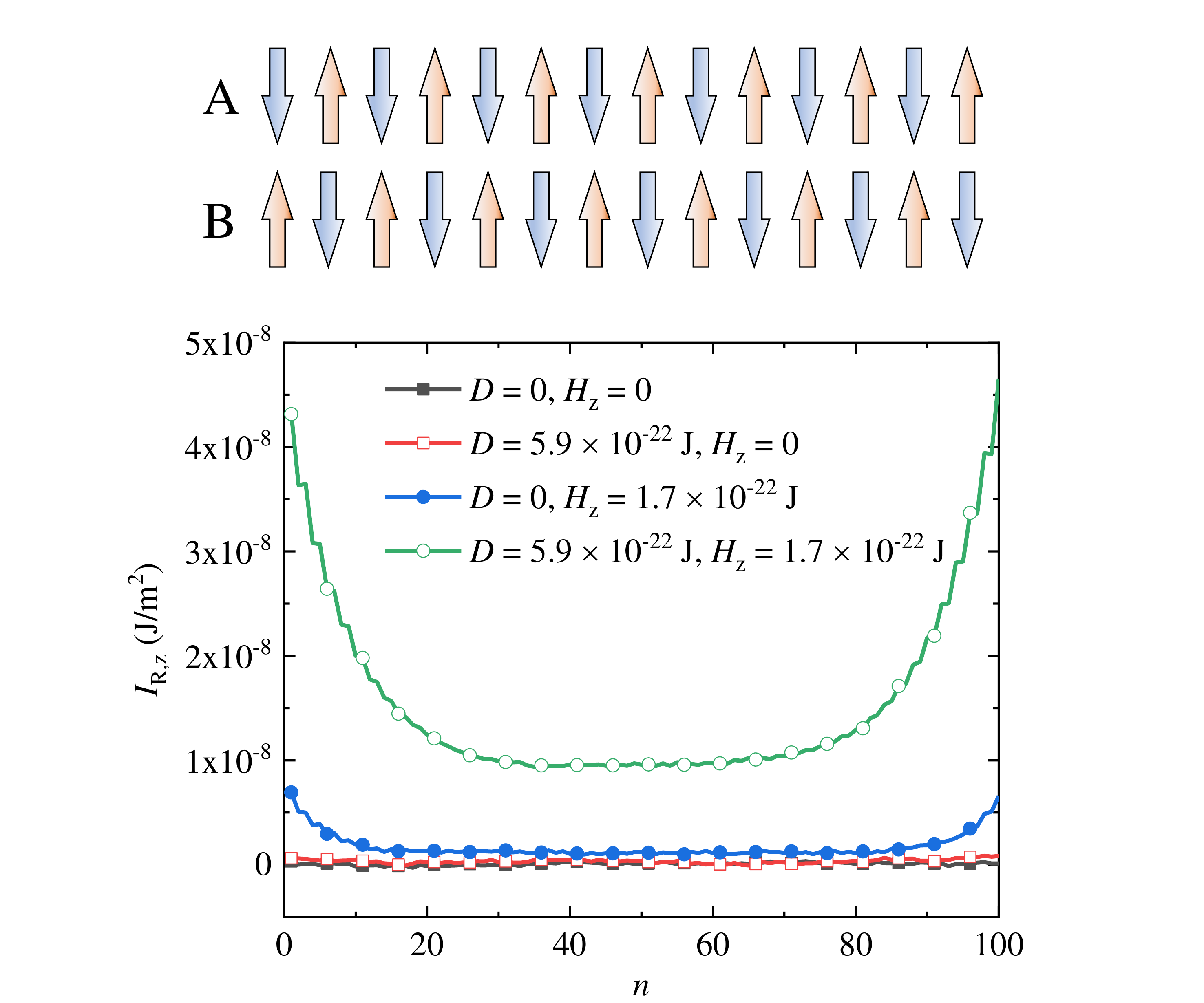}
	\caption{\label{profile-pump}  Spatial profile of the pumping current $ I_R $. The pumping current is determined through  numerical calculations  using  Eq. (\ref{spincurrent}). Two types of collinear anti-parallel states are considered, and the final result is averaged over  both A and B configurations. The spin pumping current is enhanced at the edges due to the magnon accumulation.}
\end{figure}

At first, we consider a uniform temperature $ T = 30 $ K in AFM and zero temperature $ 0 $ K in the adjacent NM. Due to the temperature gradient between the AFM and NM, the thermally activated spin dynamics in the AFM  pumps a finite spin current into the NM. Its magnitude depends on the spin mixing conductance at the  AFM/NM interface. 
Here, the spin pumping current is determined numerically based on  Eq. (\ref{spincurrent}).  Small fluctuations in the current are due to the random thermal field.

In the absence of  external magnetic fields and DM interaction, the two AFM modes are degenerate, so the number of magnons with   opposite spin orientations are equal. Therefore, though the total magnon current is nonzero, the magnonic spin pumping current vanishes, as the currents corresponding to different modes compensate for each other, see Fig. \ref{profile-pump}.
 The external magnetic field itself, applied along the easy ($ z $) axis, breaks the symmetry of the two magnon modes and lifts their degeneracy. This leads to a nonzero net spin pumping current flowing along the $ z $ axis. In Fig. \ref{profile-pump}, the averaged pumping current $ I_R $ is positive for the magnetic field $ H_z = 1.7 \times 10^{-22} $ J, and it is relatively small.

The DM interaction also lifts the degeneracy of the magnon modes as well, and shifts the magnon dispersion curves horizontally. However, in the absence of a magnetic field, the mirror (left-right) symmetry in the magnonic spectrum is still preserved, see Fig. \ref{dispersion}, and the net pumping current is zero, as shown in Fig. \ref{profile-pump}. The combination of the external magnetic field $ H_z $ and the DM interaction $ D $  strongly enhances the pumping current, see Fig. \ref{profile-pump}.

\begin{figure}[htbp]
	\includegraphics[width=0.48\textwidth]{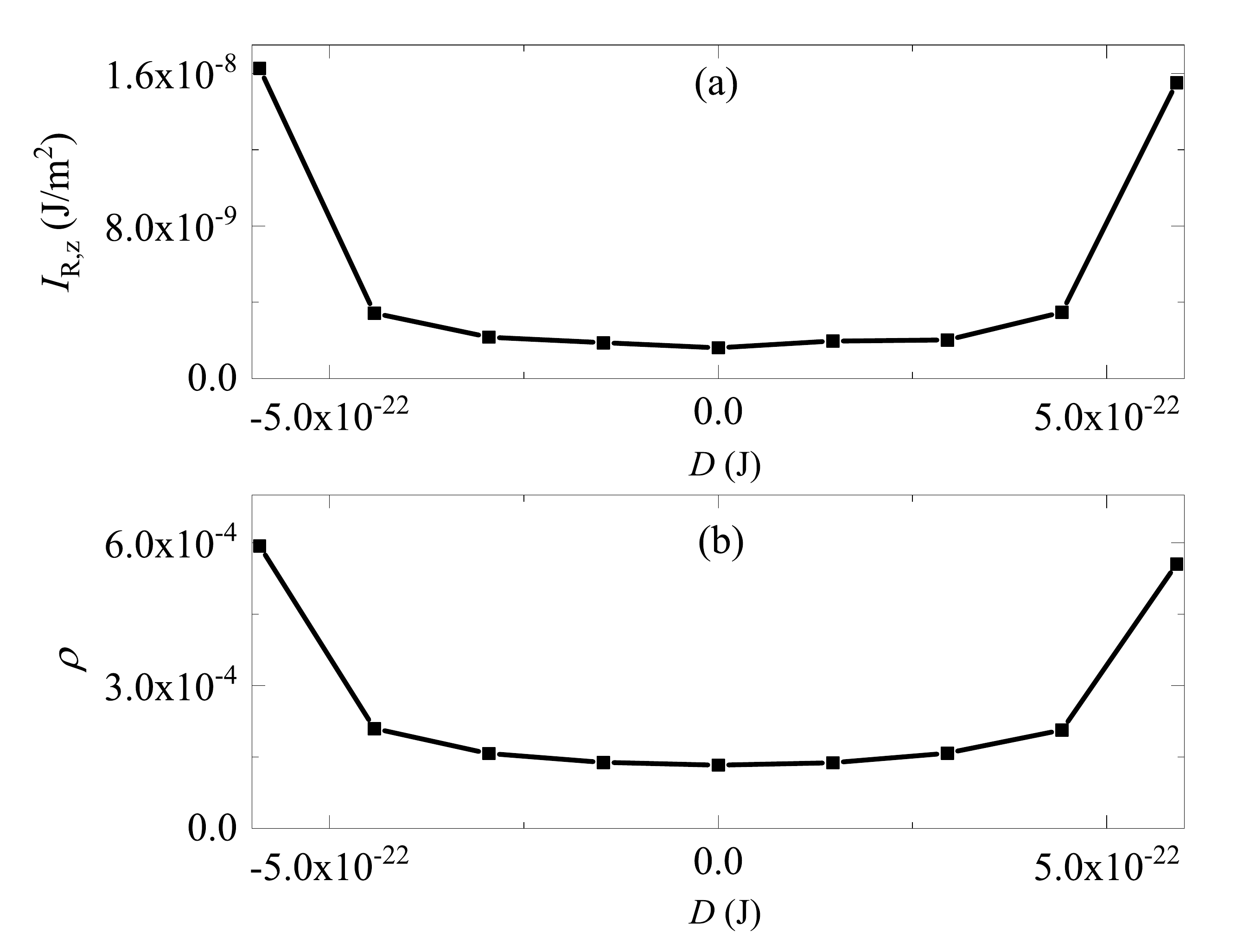}
	\caption{\label{Df}  The average pumping current $ I_R $ (a), and the magnon density $ \rho $ (b) as a function of the DM interaction strength  $ D $ for the magnetic field $ H_z = 1.7 \times 10^{-22} $ J. The antiferromagnetic collinear ground state is stable for $ |D| < \sqrt{(4I + K) K} = 9.8 \times 10^{-22} $ J.}
\end{figure}

\begin{figure}[htbp]
	\includegraphics[width=0.48\textwidth]{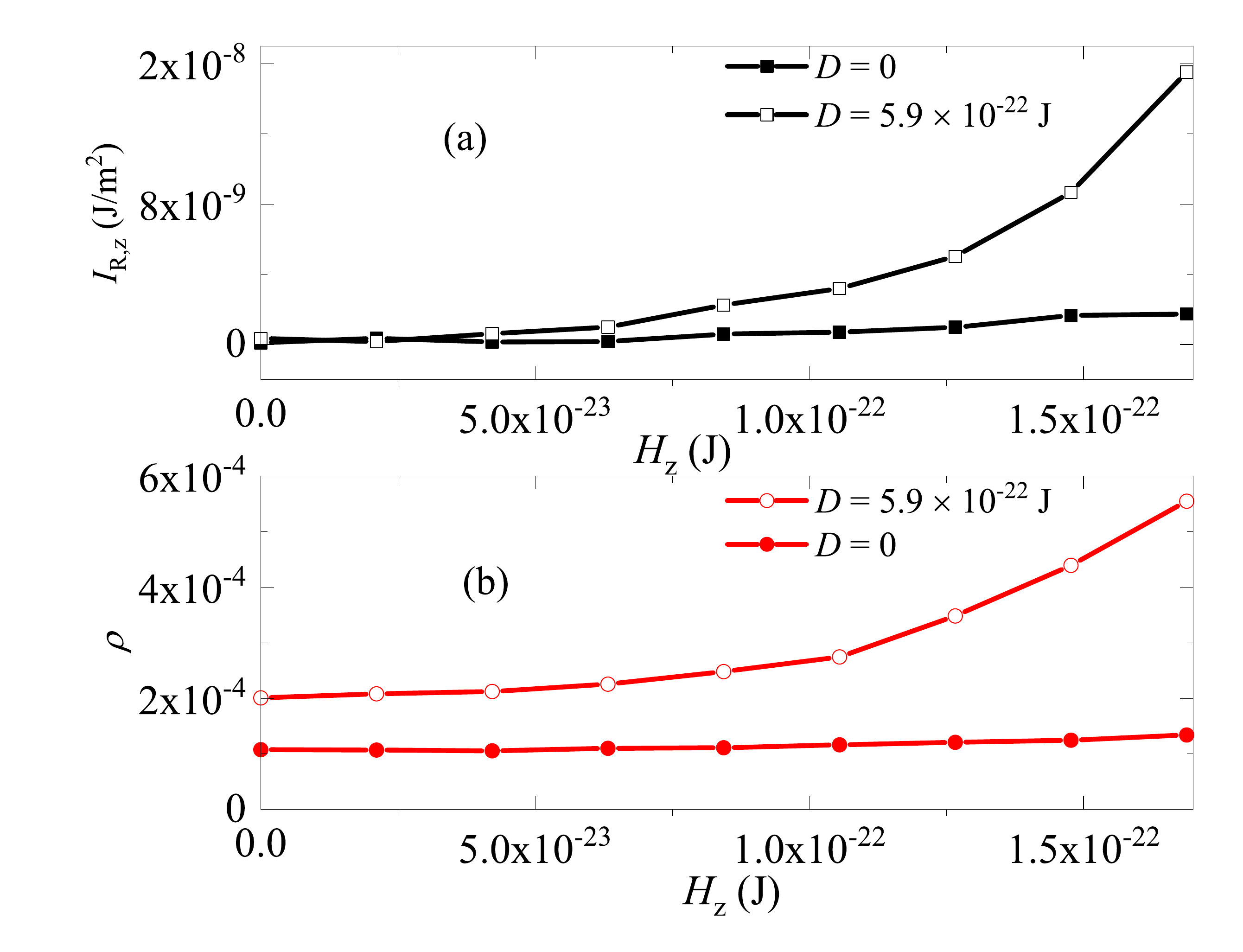}
	\caption{\label{hf}  The average pumping current $ I_R $ (a), and the magnon density $ \rho $ (b) as a function of the magnetic field $ Hz $ for $ D = 0 $, and for $ D = 5.9 \times 10^{-22} $ J. The antiferromagnetic collinear ground state is stable for $ H_z < 2S \sqrt{I K} = 1.8 \times 10^{-22} $ J.}
\end{figure}

The spin pumping current increases with the DM constant $D$. In Fig. \ref{Df} we show  this dependence of the spin pumping current for a constant magnetic field.  For small values of the DM constant D, the current increases rather slowly. However, after reaching a certain threshold value of $D$, the current increases relatively fast.  It is worth noting, that the spin pumping current  $ I_R $ is symmetric with respect  to  the sign of $D$, $ I_R (-D)=I_R (D)$. Dependence of the spin pumping current on the magnitude of magnetic field is shown in Fig. \ref{hf} for $D=0$ and also for a nonzero value of $D$. One can see again that the magnetic field itself (in the absence of DM interaction) leads to a rather small spin pumping current, while the presence of the DM interaction  increases strongly the spin current. Moreover, for large external magnetic fields one can  observe a nonlinear increase of the spin pumping current with the field.

\subsection{Spin currents in the case of a temperature gradient}

\begin{figure}[htbp]
	\includegraphics[width=0.48\textwidth]{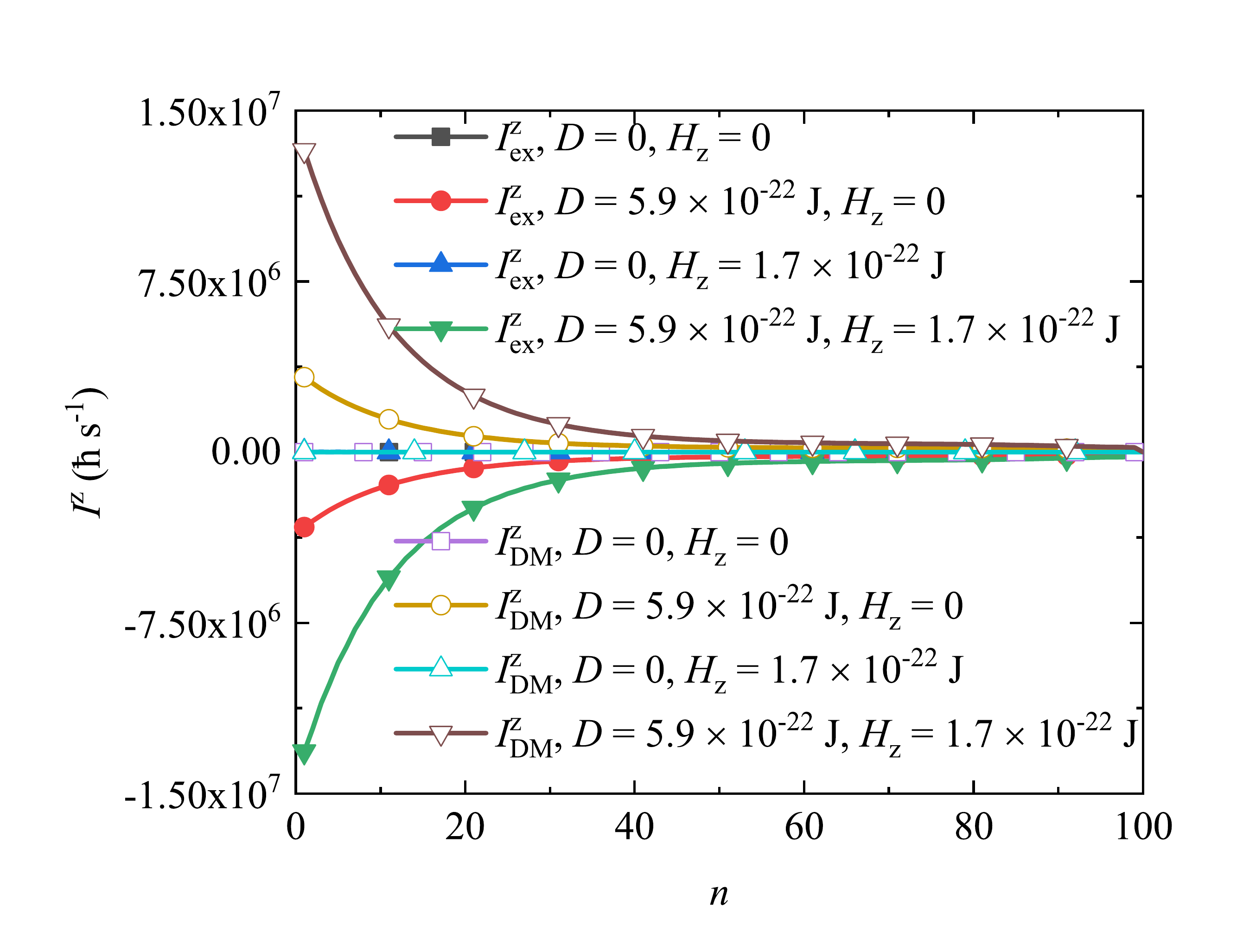}
	\caption{\label{profile_current} Spatial profiles of the exchange $ I_{ex}^z $ and DM $ I_{DM}^z $ spin currents in the case of a nonuniform temperature profile in AFM. A heat bath of the temperature $ T = 30 $ K is attached at the left edge of the AFM chain.  All generated spin currents decay away from the heat source.}
\end{figure}

\begin{figure}[htbp]
	\includegraphics[width=0.48\textwidth]{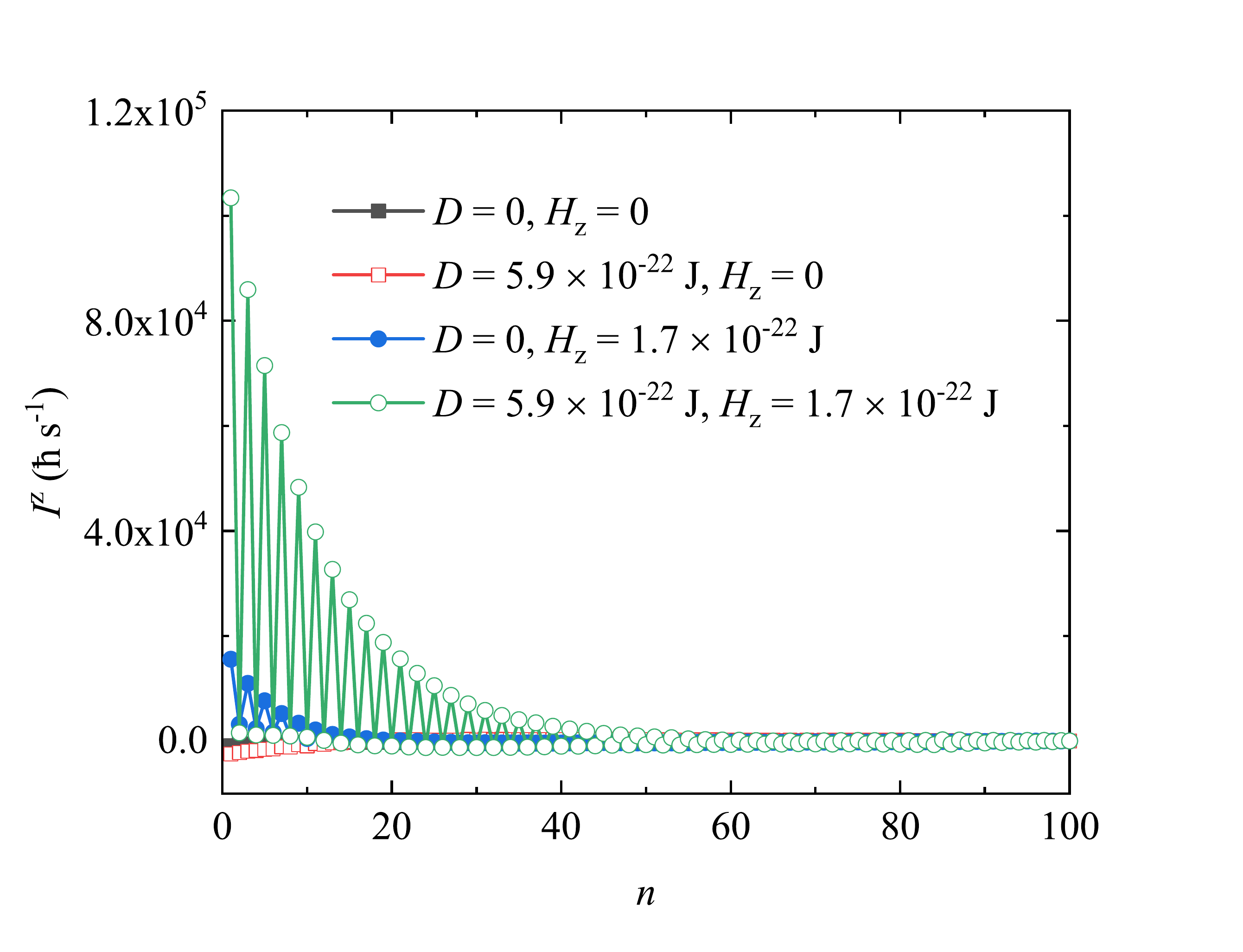}
	\caption{\label{profile_currentsum} The spatial profile of the total current $I^z= I_{ex}^z +  I_{DM}^z $, for  different sets of parameters, and in the case of nonuniform temperature profile in the AFM. The heat bath with the temperature $ T = 30 $ K is attached at the left edge of the AFM nanowire.  The total current decays with  distance away from the heat source, and also reveals oscillations with the number of sites, with the period equal to 2 sites. }
\end{figure}

\begin{figure}[htbp]
	\includegraphics[width=0.48\textwidth]{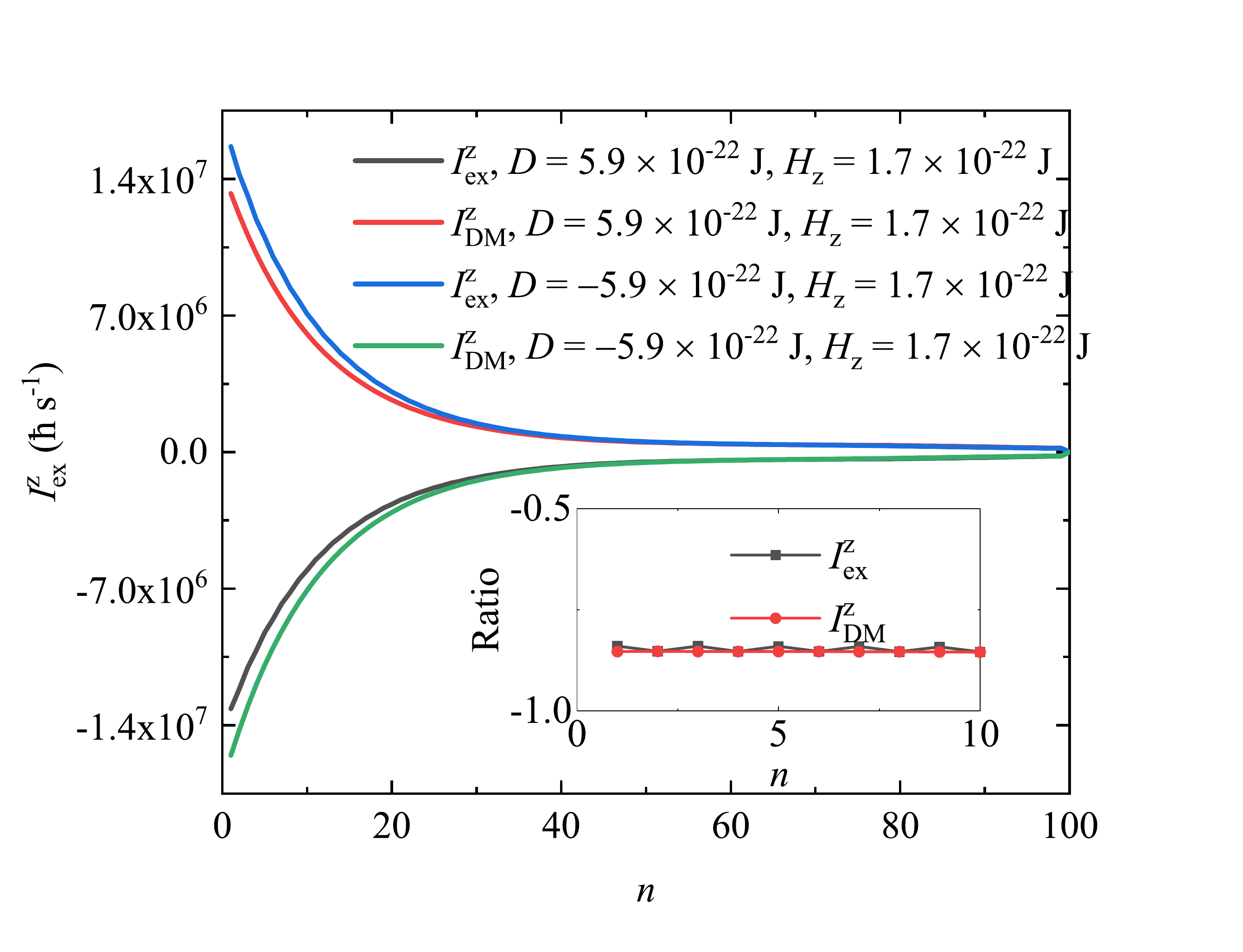}
	\caption{\label{ratio} The spatial profiles of the spin currents $ I_{ex}^z $ and $ I_{DM}^z $ for DM interaction strength  $ D = 5.9 \times 10^{-22} $ J, and $D= -5.9 \times 10^{-22} $ J, in the case of a nonuniform temperature profile in the AFM. The heat bath with a temperature $ T = 30 $ K is attached to the left edge of the AFM nanowire. The inset shows the ratio of $ I_{ex}^z $ and $ I_{DM}^z $ for positive and negative $D$. }
\end{figure}

\begin{figure}[htbp]
	\includegraphics[width=0.48\textwidth]{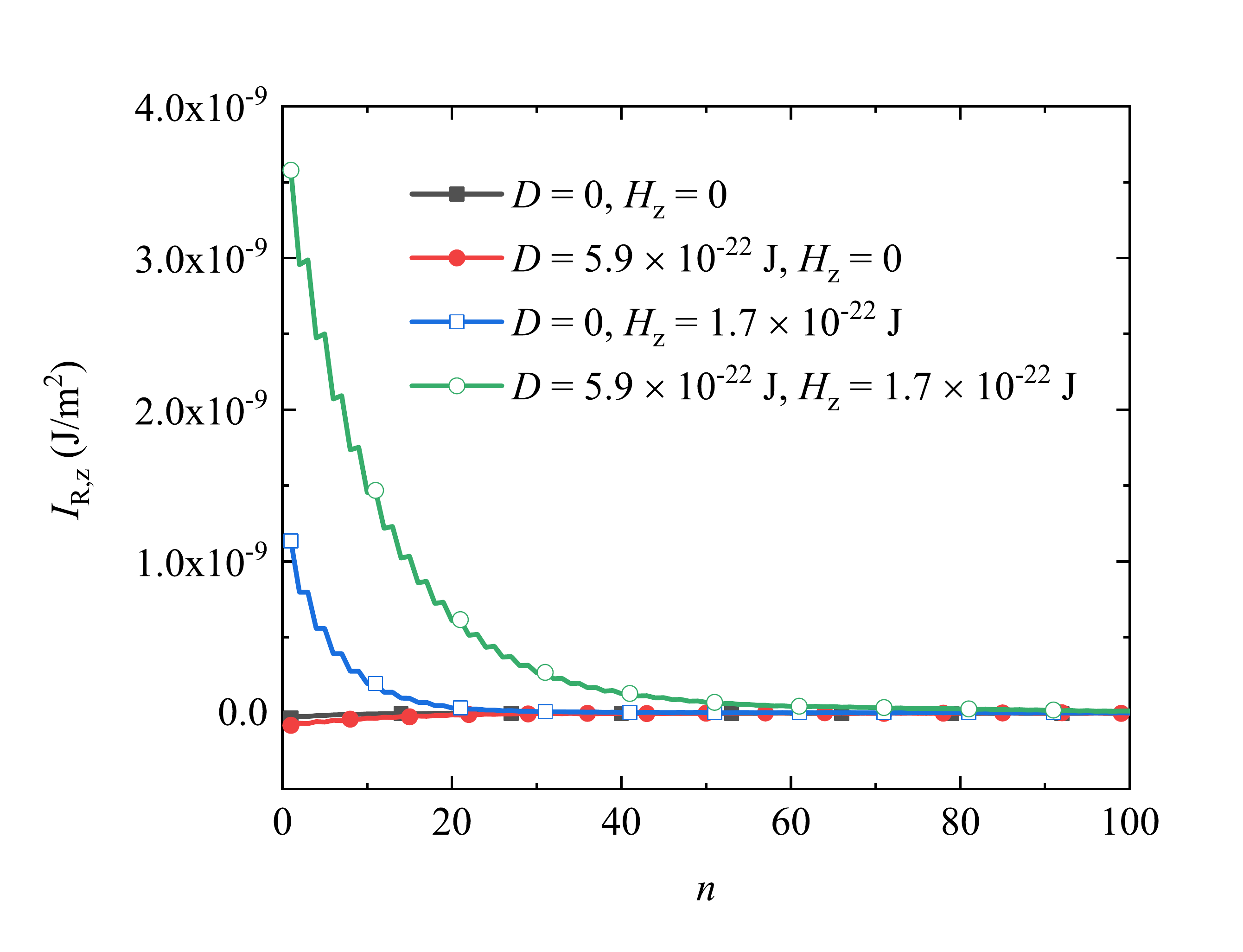}
	\caption{\label{profile_pumpg} The spatial profile of the pumping current $ I_R $
	in the case of a nonuniform temperature profile in the AFM (the heat bath with the temperature $ T = 30 $ K is attached at the left edge of the AFM nanowire). The pumping current is determined numerically using  Eq. (\ref{spincurrent}). }
\end{figure}

We set now the temperature $ T = 30 $ K at the left edge of the AFM nanowire. The temperature of the normal metal and the right edge is zero. Thus, we have two different temperature gradients: The uniform gradient profile formed in the AFM and the temperature gradient between AFM and NM. 
The first one generates exchange and DM magnonic spin currents, $I_{ex} $ and $ I_{DM}$, along the $-y$  axis, whereas the second gradient leads to the spin pumping current from the AFM to the NM.  Due to the system's geometry, the spin pumping current in our case originates from the thermally activated spin dynamics in AFM and is not related to the spatial transport of magnons (magnons flow along the $-\textbf{y}$  axis from the hot to the cold edge of AFM). Therefore, the magnon Doppler effect is not relevant for the spin pumping current, and the symmetry $I_{R} (-D)=I_{R} (D)$ is preserved. On the other hand, we expect to see an asymmetry in the exchange $I_{ex} $ and DM magnonic  $ I_{DM}$ spin currents for $\pm D$  due to the larger propagation distance of magnons along the $-\textbf{y}$ axis. The total magnonic spin current $ I^z=I_{DM} + I_{ex} $ includes  contributions from the exchange  and   DM currents.  We calculate the spin currents using Eqs. (\ref{desecrate formula}-\ref{dmicurrent}). The results are plotted in Fig. \ref{profile_current} and  Fig. \ref{profile_currentsum}. Without DM interaction and zero external magnetic field, the DM  and exchange currents vanish,  $ I_{DM}  = 0 $ and $ I_{ex} =0$.  This is because  the two AFM bands are then  degenerate.
\begin{figure}[htbp]
	\includegraphics[width=0.48\textwidth]{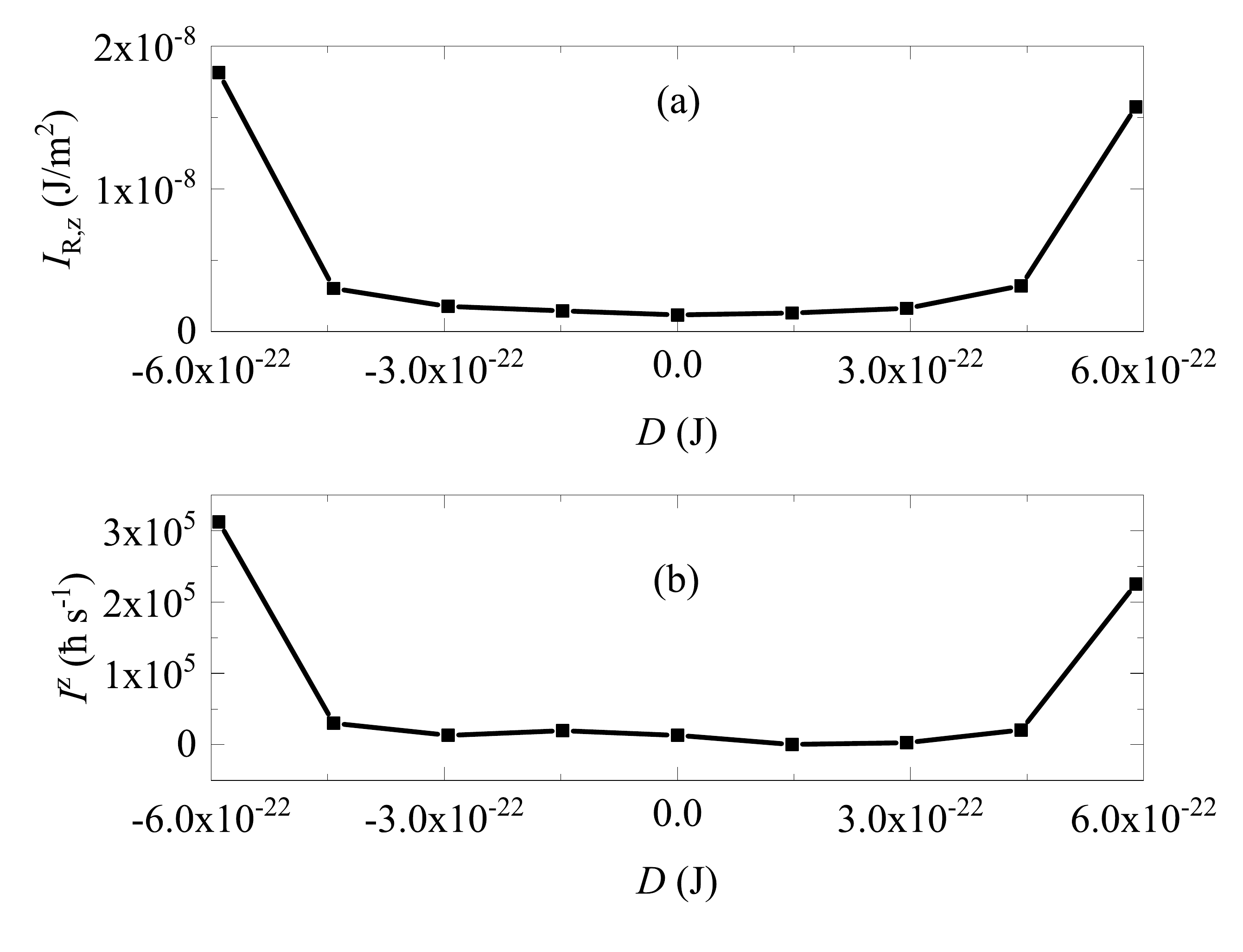}
	\caption{\label{Dfg}  The average pumping current $ I_R $ (a), and the total spin current $ I^z = I_{DM}^z + I_{ex}^z $ (b) as a function of the DM  interaction strength $ D $   for the magnetic field $ H_z = 1.7 \times 10^{-22} $ J.  The  left-right $\pm D$ asymmetry is more pronounced in the case of DM current while in the case of an exchange current the asymmetry appears only for stronger  DM interaction.}
\end{figure}

An applied magnetic field with $ H_z = 1.7 \times 10^{-22} $ J generates a positive exchange current  (see Fig. \ref{profile_currentsum}). The DM interaction leads to nonreciprocal magnons \cite{PhysRevB.98.020401,PhysRevB.101.224419,PhysRevB.96.180414},  and the exchange current $ I_{ex} $ becomes negative (Fig. \ref{profile_currentsum}).  The positive DM current $ I_{DM} $ offsets the exchange current, and the total current is zero (Fig. \ref{profile_currentsum}). Apart from this, the combination of the magnetic field $ H_z = 1.7 \times 10^{-22} $ J and the DM interaction with strength  $ D = 5.9 \times 10^{-22} $ J   enhances strongly both the  negative $ I_{ex} $ and the positive $ I_{DM} $ currents, and the positive total current is  larger than that for the case of $ D = 0 $. We observe an asymmetry between the currents generated for positive and negative $D$. For example, Fig. \ref{ratio} shows $ I_{ex}^z $ and $ I_{DM}^z $, obtained for $ D = 5.9 \times 10^{-22} $ J and $ -5.9 \times 10^{-22} $ J, and the ratio of the currents for positive and negative $D$ is smaller than 1. 

A notable  result is the fast oscillations of the total magnonic spin current, as shown in  Fig. \ref{profile_currentsum}. The period of these oscillations is equal to 2 lattice sites. This effect has the following explanation: The nonuniform magnon density profile causes the rapid oscillations of the total magnon current $ I^z $. The applied external magnetic field $ H_z $ breaks the symmetry of the effective internal fields in two sublattices leading to the spatially resolved nonuniform magnon density (Fig. \ref{profile-rou}). The spatial gradient of the magnon density generates the total magnon current. The gradient of the magnon density is not uniform, and therefore the total spin current  is also not uniform. This statement is supported by the analysis of the exchange and DM currents  $ I^z_{\rm DM} $ and $ I^z_{\rm ex} $. As we see in the inset of Fig. \ref{ratio}, the ratio between the currents is changing with  the lattice index $n$.

The spin pumping current for a nonuniform temperature profile $ I_R $ is plotted in Fig. \ref{profile_pumpg}.  One  notes that the pumping current decays away from the left edge of the AFM. The reason is that only the left edge of the AFM is in contact with the thermal bath. Therefore, the thermal magnons are generated locally and their propagation is limited by the magnon propagation length. To compare the different natures of the total spin current $ I^z=I_{DM} + I_{ex} $ and the spin pumping current $I_R $, we plot their dependence on the DM interaction strength $D$, see Fig. \ref{Dfg}. In the case of  the total spin current $I^z$, we observe a left-right ($\pm D$) asymmetry even for a small $D$, while in the case of the  pumping current $I_R $ the asymmetry is pronounced only for large $D$.

\section{Summary and conclusions}

We studied the spin Seebeck effect (SSE) in an AFM model system with DM interaction.  We  calculated  the exchange and DM spin currents in the presence of a thermal bias. In the absence of the DM interaction and an external magnetic field, there is a magnon current associated with the temperature gradient, while the corresponding spin current vanishes. This is because the contributions to spin currents that stem  from the two magnon modes are opposite. A substantial  spin current (and thus also a nonzero spin Seebeck effect) necessitates a nonzero magnetic  field and a non-vanishing  DM interaction strength. We   calculated the pumping spin current into a nonmagnetic metal  adjacent to the AFM in the presence of a thermal bias between AFM and NM as well as within the antiferromagnet.  The results highlight the important role of the DM interaction which introduces  a left-right propagation asymmetry. The derived results  supports the usefulness  of  AFM systems  for applications in spin caloritronics.

\begin{acknowledgments}
This work is supported by the DFG through  SFB-TRR 227 (B06) and project No. 465098690, by Shota Rustaveli National Science Foundation of Georgia (SRNSFG) [Grant No. FR-19-4049], the National Research Center in Poland as a research project No. DEC-2017/27/B/ST3/02881, (VKD) and within the
Norwegian Financial Mechanism 2014-2021 under the Polish-Norwegian
Research Project NCN GRIEG (2Dtronics) no. 2019/34/H/ST3/00515 (AD,JB),
and by the National Natural Science
Foundation of China (Nos.12174452, 12074437 and 11704415) and Natural Science Foundation of Hunan Province of China (Nos. 2021JJ30784 and 2020JJ4104).
\end{acknowledgments}

\begin{appendix}

\section{}

In this appendix we consider the general case, $\mathbf{D}_{ij}^{\alpha =2} = \kappa \mathbf{D}_{ij}^{\alpha =1}$, with $\mathbf{D}_{ij}^{\alpha =1}\equiv  \mathbf{D}_{ij}$,  and also present additional technical  details. Upon the Holstein and  Fourier transformations, the Hamiltonian takes on the form
\begin{eqnarray}
    \hat{H} = \sum_{\mathbf{k}} \left[ \mathcal{F}(\mathbf{k}) (\hat{a}_{\mathbf{k}}^{+} \hat{a}_{\mathbf{k}} + \hat{b}_{\mathbf{k}}^{+} \hat{b}_{\mathbf{k}}) - \mathcal{G}(\mathbf{k}) (\hat{a}_{\mathbf{k}}^{+} \hat{a}_{\mathbf{k}} +\kappa \hat{b}_{\mathbf{k}}^{+} \hat{b}_{\mathbf{k}})\right. \nonumber \\
     \left. + H_z (\hat{a}_{\mathbf{k}}^{+} \hat{a}_{\mathbf{k}} - \hat{b}_{\mathbf{k}}^{+} \hat{b}_{\mathbf{k}}) + \mathcal{C} (\hat{a}_{\mathbf{k}}^{+} \hat{b}_{- \mathbf{k}}^{+} + \hat{a}_{\mathbf{k}} \hat{b}_{- \mathbf{k}})\right],\hspace{0.5cm}
\end{eqnarray}
where
the functions $\mathcal{F}(\mathbf{k})$, $\mathcal{G}(\mathbf{k})$ and $\mathcal{C}$ are given by Eqs.(3). 
Upon the Bogolyubov transformation the Hamiltonian reads
\begin{widetext}
\begin{align}
\hat{H} = & \sum_{\mathbf{k}} \left[\mathcal{F}(\mathbf{k}) (u^{2} + \nu^{2}) - \mathcal{G}(\mathbf{k}) (u^{2} + \kappa \nu^{2}) + H_z + 2u\nu \mathcal{C} \right] \hat{c}^{+}_{\mathbf{k}} \hat{c}_{\mathbf{k}}\nonumber\\
& + \sum_{\mathbf{k}} \left[\mathcal{F}(\mathbf{k}) (u^{2} + \nu^{2}) - \mathcal{G}(\mathbf{k}) (\kappa u^{2} + \nu^{2}) - H_z + 2 u\nu \mathcal{C} \right] \hat{d}^{+}_{\mathbf{k}} \hat{d}_{\mathbf{k}}\nonumber\\
&+ \sum_{\mathbf{k}} \left[2 u \nu \mathcal{F}(\mathbf{k}) - u \nu (1+\kappa) \mathcal{G}(\mathbf{k}) + \mathcal{C} (u^{2} + \nu^{2}) \right] (\hat{c}_{\mathbf{k}}^{+} \hat{d}_{-\mathbf{k}}^{+} + \hat{c}_{\mathbf{k}} \hat{d}_{-\mathbf{k}}).
\end{align}
\end{widetext}
The above Hamiltonian becomes diagonal when the following condition is fulfilled: 
\begin{equation}
2 u \nu \mathcal{F}(\mathbf{k}) - u \nu (1+\kappa) \mathcal{G}(\mathbf{k}) + \mathcal{C} (u^{2} + \nu^{2}) = 0,
\end{equation}
or equivalently
\begin{equation}
2 u \nu \mathcal{W}(\mathbf{k}) + \mathcal{C} (u^{2} + \nu^{2}) = 0  ,
\end{equation}
where
\begin{equation}
    \mathcal{W}(\mathbf{k}) = \mathcal{F}(\mathbf{k}) - \frac{1 +\kappa}{2} \mathcal{G}(\mathbf{k}).
\end{equation}
Equation (A3) and the condition $u^2-\nu^2 =1$ give the following solution for $u$ and $\nu$:
\begin{eqnarray}
\left\{\begin{array}{l}
\nu^{2} = \frac{1}{2} \frac{\mathcal{W}(\mathbf{k})}{\sqrt{\mathcal{W}(\mathbf{k})} - \mathcal{C}^{2}} - \frac{1}{2}  \\
u^{2} = \frac{1}{2} \frac{\mathcal{W}(\mathbf{k})}{\sqrt{\mathcal{W}(\mathbf{k})} - \mathcal{C}^{2}} + \frac{1}{2} .
\end{array}\right. 
\end{eqnarray}
The eigenvalues have  then the  following form:
\begin{equation}
    \varepsilon_{1,2} = \sqrt{\left[\mathcal{F}(\mathbf{k}) - \frac{1+\kappa}{2} \mathcal{G}(\mathbf{k})\right]^{2} - \mathcal{C}^{2}} \mp \frac{1-\kappa}{2} \mathcal{G}(\mathbf{k}) \pm H_z.
\end{equation}
Taking into account the definitions (10),  we find  the following formula for the eigen-energies (up to the linear terms in $\varepsilon_{DM}$):
\begin{eqnarray}
   \varepsilon_{\mathbf{k}1,2} = \sqrt{\varepsilon_{exA}(\varepsilon_{exA}+2\varepsilon_{AF})} \mp \frac{1-\kappa}{2}\varepsilon_{DM}\nonumber \\
    - \frac{1+\kappa}{2}\frac{\varepsilon_{exA}+\varepsilon_{AF}}{\sqrt{\varepsilon_{exA}(\varepsilon_{exA}+2\varepsilon_{AF})}}\varepsilon_{DM}\pm H_z.
\end{eqnarray}
In the cases of $\kappa =1$ and $\kappa =-1$, this formula reduces to Eqs (9) and (12), respectively.
When the anisotropy dominates over the other interactions, one may write the eigenenergies as
\begin{equation}
    \varepsilon_{\mathbf{k}1,2}
     \approx \varepsilon_{ExA} + \varepsilon_{AF} - \frac{1+\kappa \pm (1-\kappa )}{2}\varepsilon_{DM} \pm H_z,
\end{equation}
which reduces to (11) and (13) for  $\kappa =1$ and $\kappa =-1$, respectively
\end{appendix}
\newpage
\bibliographystyle{apsrev4-1}
\bibliography{DM-AFM}
\end{document}